\documentstyle[epsfig]{mn}
\begin{document}
\def\ltsima{$\; \buildrel < \over \sim \;$}
\def\simlt{\lower.5ex\hbox{\ltsima}}
\def\gtsima{$\; \buildrel > \over \sim \;$}
\def\simgt{\lower.5ex\hbox{\gtsima}}
\def\approxgt{\mathrel{\hbox{\rlap{\lower.55ex \hbox {$\sim$}}
        \kern-.3em \raise.4ex \hbox{$>$}}}}
\def\approxlt{\mathrel{\hbox{\rlap{\lower.55ex \hbox {$\sim$}}
        \kern-.3em \raise.4ex \hbox{$<$}}}}

\title{Warm Absorber Energetics in Broad Line Radio Galaxies}

\author[E.Torresi et al.]
{E.~Torresi$^{1,2}$, P.~Grandi$^{1}$, E.~Costantini$^{3}$, G.G.C.~Palumbo$^{2}$\\
$^{1}$Istituto di Astrofisica Spaziale e Fisica Cosmica-Bologna, INAF, 
via Gobetti 101, I-40129 Bologna \\
$^{2}$Dipartimento di Astronomia, Universit\`a di Bologna, via Ranzani 1, I-40127 Bologna, Italy \\
$^{3}$SRON, Netherlands Institute for Space Research, Sorbonnelaan 2, 3584 CA Utrecht, The Netherlands\\
}

\date{Accepted 2011 August 24. Received in original form 2011 March 7}

\maketitle

\begin{abstract}
We review the soft X--ray properties of 3C~390.3, 3C~120, 3C~382 and 3C~445, the only  Broad Line Radio Galaxies (BLRG) for which good  quality gratings data are currently available.  The \textit{XMM--Newton}/RGS data of 3C~390.3 and 3C~120 were re--analyzed searching for warm absorbers, already discovered in 3C~382 and 3C~445.
We confirm the absence of  ionized absorption features  in 3C~120, but find signatures of outflowing gas (v$_{out} \sim$ 10$^{2}$~km~s$^{-1}$) in 3C~390.3.
Its warm absorber (Log$\xi \sim$ 2~erg~cm~s$^{-1}$, N$_{H}\sim$ 10$^{20}$~cm$^{-2}$), similar to that observed in 3C~382,  is  probably placed in the Narrow Line Regions. Its gas content is slower and less dense than the accretion disk wind discovered in 3C~445. 
Independently from the location of the warm gas, the outflowing masses ($\dot{M}_{out}$) of BLRGs are significantly  (but improbably) predominant with respect to the accretion masses ($\dot{M}_{acc}$),  suggesting a clumpy configuration of the warm absorber. However, even assuming overestimated values of $\dot{M}_{out}$,  the kinetic luminosity of the outflow ($\dot{E}_{out}$) is well below 1$\%$ of the kinetic power of the jet (P$_{jet}$). Thus, the jet remains the major driver of the radio--loud AGN feedback at least on pc scale and beyond.
The warm absorber parameters (N$_{H}$, $\xi$) of BLRGs span similar range of values of type 1 radio--quiet AGNs.
However, when the mass outflow rate of BLRGs and Seyfert 1s is plotted as a function of the radio--loudness, R=Log [$\nu$ L$_{\nu(5GHz)}$/L$_{(2-10keV)}]$, the mass outflow rate seems to increase with radio power.
\end{abstract}


\vspace{1.0cm}

\section{Introduction}
In the last few years high--resolution X--ray spectroscopy has made progress in the exploration of the circumnuclear environment of radio--loud (RL) AGNs.
Studies performed on obscured RL sources (Grandi et al. 2007 hereafter G07; Sambruna et al. 2007; Reeves et al. 2010 hereafter R10; Piconcelli et al. 2008; Torresi et al. 2009; Evans et al. 2010) revealed photoionized emitting--line gas as responsible for the soft--excess, similarly to radio--quiet (RQ) Seyfert  2 galaxies.
If absorption and emission are processes occurring in the same plasma, as suggested for Seyfert galaxies (Kinkhabwala et al. 2002), it  is natural to assume that warm absorbers (WA), characterizing at least 50$\%$ of Seyfert 1 spectra,  should be present also in unobscured Broad Line Radio Galaxies (BLRG). 
With the term warm absorber we intend ionized outflowing gas in our line--of--sight that produces narrow absorption lines of several elements from C to Fe in the soft X--ray spectrum. Generally, these structures are blueshifted with moderate velocities, v$\sim$100-1000~km~s$^{-1}$ (Kaastra et al. 2000; Kaspi et al. 2002; Crenshaw et al. 2003), but can reach also v$>$10$^{4}$~km~s$^{-1}$ (Pounds et al. 2003; Reeves et al. 2003, 2010; Braito et al. 2011) when the wind originates directly from the disk.
The detection of WAs in BLRGs was expected to be more difficult than in Seyfert 1s, because of the jet. If it is closer to the line--of--sight, the Doppler--boosted, non--thermal radiation could mask the absorption features.\\
Hints of WAs have been observed in the past in a handful of RL sources: 3C~382 and 3C~390.3 with  \textit{ASCA} (Reynolds 1997); 3C~351 with \textit{ROSAT} PSPC (Nicastro et al. 1999); 4C+74.26 with the \textit{XMM--Newton} EPIC cameras (Ballantyne 2005).\\
The presence of absorption features in RL AGNs was confirmed by the high--resolution X--ray analysis of 3C~382 (Torresi et al. 2010; Reeves et al. 2009). Successively R10 reported the presence of a warm absorber also in 3C~445.
Since the gas has non--negligible outflowing velocities, it could contribute in transferring momentum to the environment in addition to the jet. However to what extent is the wind importance in the energetic budget of powerful RL AGNs is still an open question.
In order to shed some light on this issue, we collect high--resolution data from literature and explore the \textit{XMM--Newton} Reflection Grating Spectrometer (RGS) archive.
The main goal is to enlarge the sample of RL sources with clear detection of warm absorbers. Finally, we attempt a RL/RQ comparison of the ionization and kinematic properties of the X--ray absorbing gas.\\
The paper is organized as follows: in Section 2 the very small sample of BLRGs considered throughout the work is described; the RGS spectral analysis performed on 3C~390.3 and 3C~120 together with the results are reported in Section 3. In Section 4 we discuss the physical and energetical properties of BLRG WAs and attempt a first comparison between RL and RQ outflows. The main results of this work are summarized in Section 5.
The cosmological values H$_{0}$=71~km~s$^{-1}$~Mpc$^{-1}$, $\Omega_{m}$=0.27, $\Omega_{\Lambda}$=0.73 (Komatsu et al. 2009) are assumed throughout. \\

\section{The sample}
The sample of BLRGs consists of four sources: 3C~390.3, 3C~120, 3C~382 and 3C~445. The main properties of each object are reported in Table~1, where the redshifts, the jet inclination angles, the black hole masses, the ionizing luminosities between 1---1000 Ry, and the radio luminosities at 151 MHz are listed.


\begin{table*}
\caption{Summary of the BLRG properties.}
\label{tab1}                
\begin{tabular}{l c c c c c}
\\
\hline                    
                                     &z  &\textit{i}   &LogM$_{BH}$$^{a}$     &LogL$_{ion}$$^{b}$        &LogL$_{151MHz}$$^{c}$   \\                           
                                      &    & (degrees)     &(M$_{\odot}$)      &(erg~s$^{-1}$)      &(W~Hz$^{-1}$~sr$^{-1}$)    \\                
\hline 
\large\textit{3C~445}  & 0.05623    &$\leq$60$^{d}$  &8.33 &44.47     &25.23   \\
\hline
\large\textit{3C~390.3} & 0.0561 &30--35$^{e}$ &8.55 &44.85 &25.53                \\
\hline
\large\textit{3C~382} &0.0579 &35--45$^{e}$ &9.06  &45.27 &25.17                  \\
\hline
\large\textit{3C~120}  &0.033 &$\leq$21$^{f}$ &7.74 &44.96 &25.04                 \\
\hline
\\
\multicolumn{6}{l}{(a) Provided by Grandi et al. (2006), with the exception of 3C~120}\\
\multicolumn{6}{l}{taken from Peterson et al. (2004).}\\
\multicolumn{6}{l}{(b) L$_{ion}$ directly measured from the proper SED of each source, except for 3C~445}\\
\multicolumn{6}{l}{taken from R10. L$_{ion}$ is the ionizing luminosity between 1--1000 Ry}\\
\multicolumn{6}{l}{($\equiv$13.6 eV--13.6 keV).}\\
\multicolumn{6}{l}{(c) L$_{151MHz}$ extrapolated from the power at 178 MHz of Hardcastle}\\
\multicolumn{6}{l}{et al. (1998) and rescaled to our $\Lambda$CDM cosmology. For 3C~120 we}\\
\multicolumn{6}{l}{refer to Arshakian et al. (2010).}\\
\multicolumn{6}{l}{(d) Estimate from the radio jet/counterjet ratio (G07).}\\
\multicolumn{6}{l}{(e) Estimate from the radio band (Giovannini et al. 2001).}\\
\multicolumn{6}{l}{(f) Upper limit on the inclination angle obtained by using the larger apparent transverse}\\
\multicolumn{6}{l}{velocity v=5.3c (Gomez et al. 2001).}\\
\end{tabular}
\end{table*}

3C~390.3 (z=0.0561; Hewitt \& Burbidge 1991) is a classical double--lobed FRII Radio Galaxy (Pearson \& Readhead 1988). It is one of the closest radio sources whose core exhibits superluminal motion in the pc--scale jet (Alef et al. 1996). From the apparent velocity of 3.5c and the core dominance, Giovannini et al. (2001) estimated the jet inclination angle $30^{\circ}<\theta<35^{\circ}$ with $\beta \sim$0.96--0.99 \footnote{$\beta$=v/c is the bulk velocity in units of the speed of light (Urry $\&$ Padovani 1995).}.
Double--peaked emission lines characterize the optical and UV spectra of the source (Eracleous \& Halpern 1994; Zheng 1996; Wamsteker et al. 1997), while the UV bump is weak or even absent (Wamsteker et al. 1997).
3C~390.3 is known to be variable in the X--ray band, with variations in both soft and hard band on a timescale of weeks to months (Leighly \& O'Brien 1997; Gliozzi et al. 2003). All previous X--ray telescopes observed this source. While in the hard X--ray band there is a general agreement on the presence of an iron line and reflection hump (Grandi et al. 1999; Sambruna et al. 2009), the modeling of the soft X--ray band is controversial. \textit{Einstein--\small{IPC}} (Kruper et al. 1990) and  \textit{BeppoSAX} (Grandi et al. 1999) required  an excess of column density.  \textit{EXOSAT} claimed the presence of a soft excess (Ghosh \& Sondararajaperumal 1991) while Reynolds (1997) found hints of warm absorption in the \textit{ASCA} data, successively confirmed by the detection of an absorption edge at 0.65 keV (Sambruna et al. 1999). Recently Sambruna et al. (2009) observed an emission line in the RGS spectrum associated with O{\small VII} forbidden {line} possibly produced in the NLR.

3C~120 (z=0.033; Burbidge 1967) is classified as an FRI exhibiting a one--sided jet (Seielstad et al. 1979; Walker et al. 1987; Harris et al. 2004). The apparent transverse velocity of the jet $v_{app}$=5.3c, as obtained by the VLBA observation (Lister et al. 2009), implies an upper limit on the inclination angle of 21$^{\circ}$. 
The optical spectrum of 3C~120 is typical of Seyfert 1 galaxies, with strong and broad emission lines, quite unusual for FRI radio sources.
Reverberation mapping constrains the black hole mass to be 5.5$^{+3.1}_{-2.3}$$\times$10$^{7}$ M$_{\odot}$ (Peterson et al. 2004).
At UV wavelengths, 3C~120 has a typical AGN spectrum with a strong blue bump and strong emission lines signature of a standard optically--thick geometrically--thin accretion disk (Maraschi et al. 1991). 
The hard X--ray spectrum is characterized by a slightly broadened iron line (EW$\sim$100 eV) at 6.4 keV, a weak ionized line at 6.9 keV (Yaqoob \& Padmanabhan 2004; Kataoka et al. 2007) and Compton reflection $\Omega$/2$\pi$$\sim$0.4--0.5 (Eracleous et al. 2000; Zdziarski \& Grandi 2001; Gliozzi et al. 2003; Ballantyne et al. 2004). The soft X--ray band of 3C~120 has been observed by the grating instruments onboard \textit{Chandra} and \textit{XMM--Newton}.  
In the \textit{Chandra} High Energy Transmission Gratings (HETG) spectrum an O{\small VIII} Ly$\alpha$ absorption line, blueshifted by $\sim$ -5500 km~s$^{-1}$, was observed (McKernan et al. 2003).   
This feature was not revealed in the \textit{XMM--Newton}/RGS observation (Ogle et al. 2005), that on the contrary shows a slightly redshifted O{\small VIII} Ly$\alpha$ emission structure. 

3C~382 (z=0.0579; Marzke et al. 1996) is a FRII lobe-dominated Radio Galaxy showing a long jet (1.68' from the core) and two radio lobes, with a total extension of 3' (Black et al. 1992).
In the optical--UV and X--ray regimes there are hints of no strong jet contamination. The optical spectrum shows broad lines (FWZI$>$25,000 km~s$^{-1}$) which are variable on a timescale of months to years. Yee and Oke (1981) suggested the presence of the UV bump from the accretion disk. 
In the X--ray band 3C~382 is a bright source (F$_{2-10 keV } \sim$ 3$\times$10$^{-11}$~erg~cm$^{-2}$~s$^{-1}$). It can be well fitted with a single power law between 2--10 keV, but shows a strong excess at lower energies (Prieto 2000; Grandi et al. 2001). In part this is probably related to an extended emission (0.2--2.4 keV) revealed by \textit{ROSAT/HRI} and \textit{Chandra} (Prieto 2000; Gliozzi et al. 2007). The presence of a slow highly ionized outflow in this source  was attested by Torresi et al. (2010) using \textit{XMM--Newton}/RGS data, and confirmed by the \textit{Chandra}/HETG spectrum (Reeves et al. 2009).

3C~445 (z=0.05623) is a powerful FRII Radio Galaxy, classified as a BLRG because of its broad and intense Balmer lines in the optical spectrum. Near--IR observations show a substantial reddening of E$_{B-V}$=1~mag and the radio--to--IR SED indicates a predominance of dust emission with a negligible contribution of synchrotron photons. The existence of absorbing material is also supported by X--ray data showing a strong depletion of the continuum photons below a few keV. R10 suggests that the nuclear view could be obscured by an outflowing and clumpy accretion disk wind with high column density (N$_{H}\sim$10$^{23}$~cm$^{-2}$) and velocity (v=10$^{4}$~km~s$^{-1}$).
In spite of its optical classification, 3C~445 seems more similar to Seyfert 2s than Seyfert 1s (G07; Sambruna et al. 2007). In agreement with that, the soft X--ray excess can be fitted with a mix of emission lines and scattered continuum produced by photoionized gas (G07; R10). Indeed, the presence of the WA in this source was deduced by the detection of a high energy absorption feature around 7 keV.\\

\noindent
Since WAs have been already ascertained in 3C~382 and 3C~445, here we re--analyze the RGS data of 3C~390.3 and 3C~120.

\section{RGS spectral analysis and results}
3C~390.3 was observed by \textit{XMM--Newton}/RGS (den Herder et al. 2001) on 2004 October 8--9 for a total exposure of 50 ks and on October 17 for 20 ks.  \\
3C~120 was pointed twice, on 2002 September 6 for 12 ks and on 2003 August 26 for 130 ks. In this paper we consider only the second and longer observation.\\
The RGS1 and RGS2 spectra were extracted using the {\small SAS} (v. 9.0.0) task {\it rgsproc}, which combines the event lists from all {\small RGS CCD}, produces source and background spectra using a region spatially offset from that containing the source, and generates response matrices. 
The resulting spectra were analyzed using the fitting package  {\small SPEX} (v.2.0) (Kaastra, Mewe and Nieuwenhuijzen 1996). The Galactic absorption was modeled with the {\small SPEX HOT} component. For all spectral models Solar elemental abundances were adopted (Anders \& Grevesse 1989).
For each source the proper line--of--sight Galactic column density was considered, N$_{H}$=3.5$\times$10$^{20}$~cm$^{-2}$ and N$_{H}$=1.1$\times$10$^{21}$~cm$^{-2}$ for 3C~390.3 and 3C~120, respectively (Kalberla et al. 2005).
The absorption/emission features were searched for following two steps:
\begin{enumerate} 
\item a phenomenological approach, consisting in the inspection of the (data--model/error) residuals after having fitted the continuum;
\item a physical approach, fitting the absorption features with the \textit{xabs} model in {\small SPEX}.
 \textit{xabs} calculates the transmission through a gas layer. Free parameters in this model are the outflow velocity ($v_{\rm out}$), the total hydrogen column density ($N_{\rm H}$) and the ionization parameter $\xi$ \footnote{$\xi$=$\displaystyle\frac{L}{n_{e}R^{2}}$, \textit{L} is the 1-1000 Rydberg (Ry) source ionizing luminosity (corresponding to 13.6 eV--13.6 keV), \textit{n$_{e}$} is the electron density of the gas and \textit{R} is the distance of the gas from the central source.}.   
In the model the column densities of different ions are linked through an ionization balance, which is precalculated using {\small CLOUDY} (Ferland et al. 1998). The ionization balance is dependent on the spectral energy distribution (SED) of the source. The UV/X--ray SEDs for both 3C~120 and 3C~390.3 were constructed using both the EPIC--pn (Str\"uder et al. 2001) and the optical monitor (OM; Mason et al. 2001) and are shown in Fig.~1.
 For  3C~390.3 we considered only the longest observation performed on Oct.~8. 
 The data were reduced using the {\small SAS} (v. 9.0.0) with standard procedures. The light curve over 10 keV was extracted to check high background periods. The source and the background spectra were extracted from circular regions of 35$''$ radius. Backgrounds were taken from a region within the same {\small CCD} of the targets and not contaminated by the sources.  The response matrices were created using the {\small SAS} commands {\small RMFGEN} and {\small ARFGEN}.
Events outside the 0.4--10 keV band were discarded in the pn spectra of both sources (Guainazzi 2010).
We produced a rough representation of the continuum using a simple absorbed power law for 3C~390.3 ($\Gamma \sim$1.8) and a broken power law ($\Gamma_1=2.04$, $\Gamma_2=1.75$ with a break at 2.5 keV) for 3C~120. For the purpose of constructing the SED, both continua were then unabsorbed, to obtain the true ionizing X--ray flux. 
For the optical/UV part of the SED we used the OM measurements for both 3C~120 (V and UVW1 filters) and 3C~390.3 (U, UVW1, UVM2 and UVW2 filters). The photometric analysis was performed using the standard procedure within the \textit{XMM--Newton} {\small SAS} (v. 9.0.0).
The optical fluxes have been dereddened using the extinction curves of  Cardelli et al. (1989), knowing the optical extinction $A_V$. This has been calculated following Bohlin et al. (1978) formula: $N_{\rm H}\sim A_V*1.9\times10^{21}$\,cm$^{-2}$ (for $R_V=3.1$). 
The low energy tail of the SED (infra--red to radio) was taken as described in the standard SED used in {\small CLOUDY} (Mathews \& Ferland 1987).   
\end{enumerate}


\begin{figure}
\begin{center}
\epsfig{file=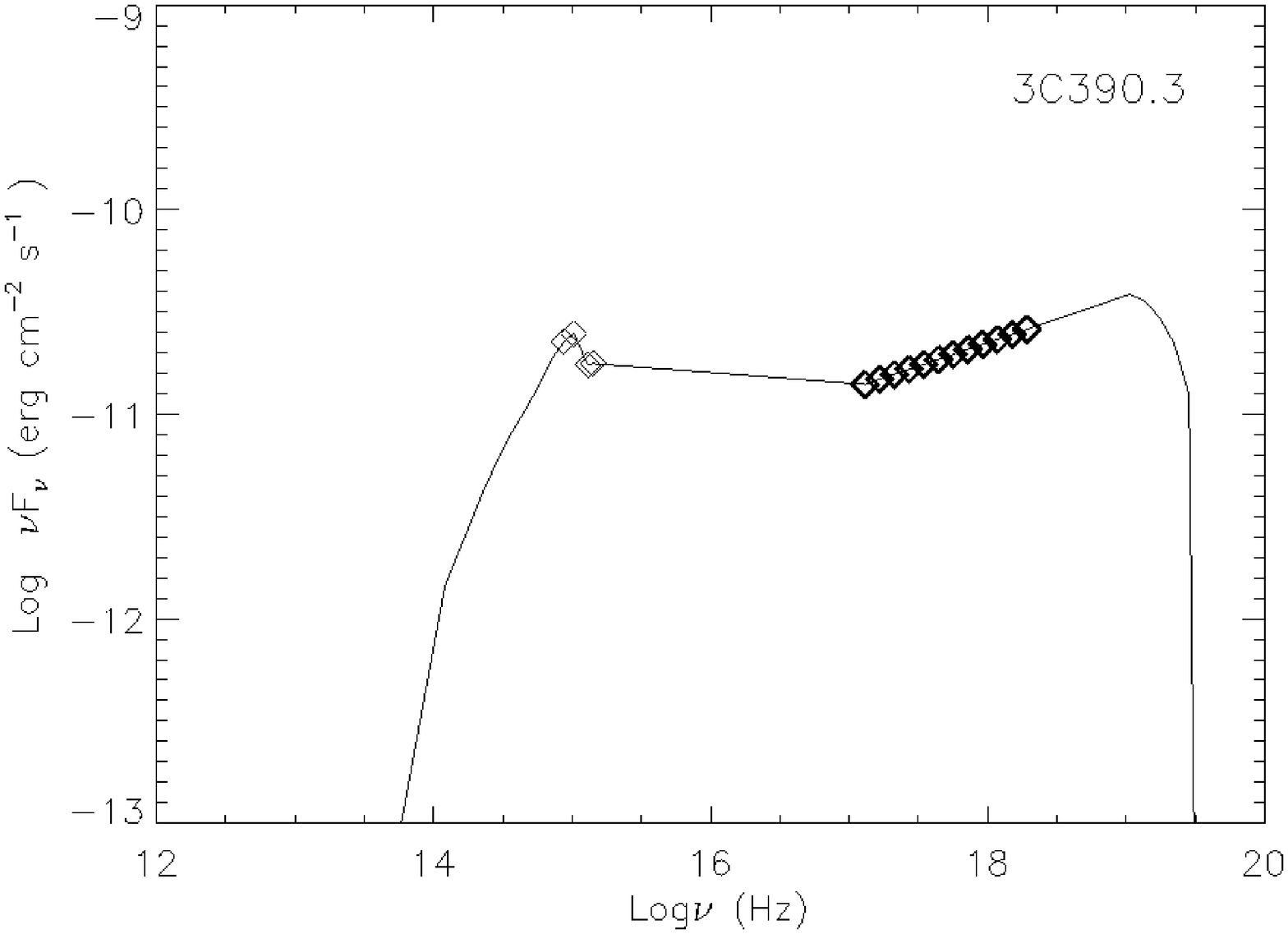,height=5cm,width=5cm, angle=90}
\epsfig{file=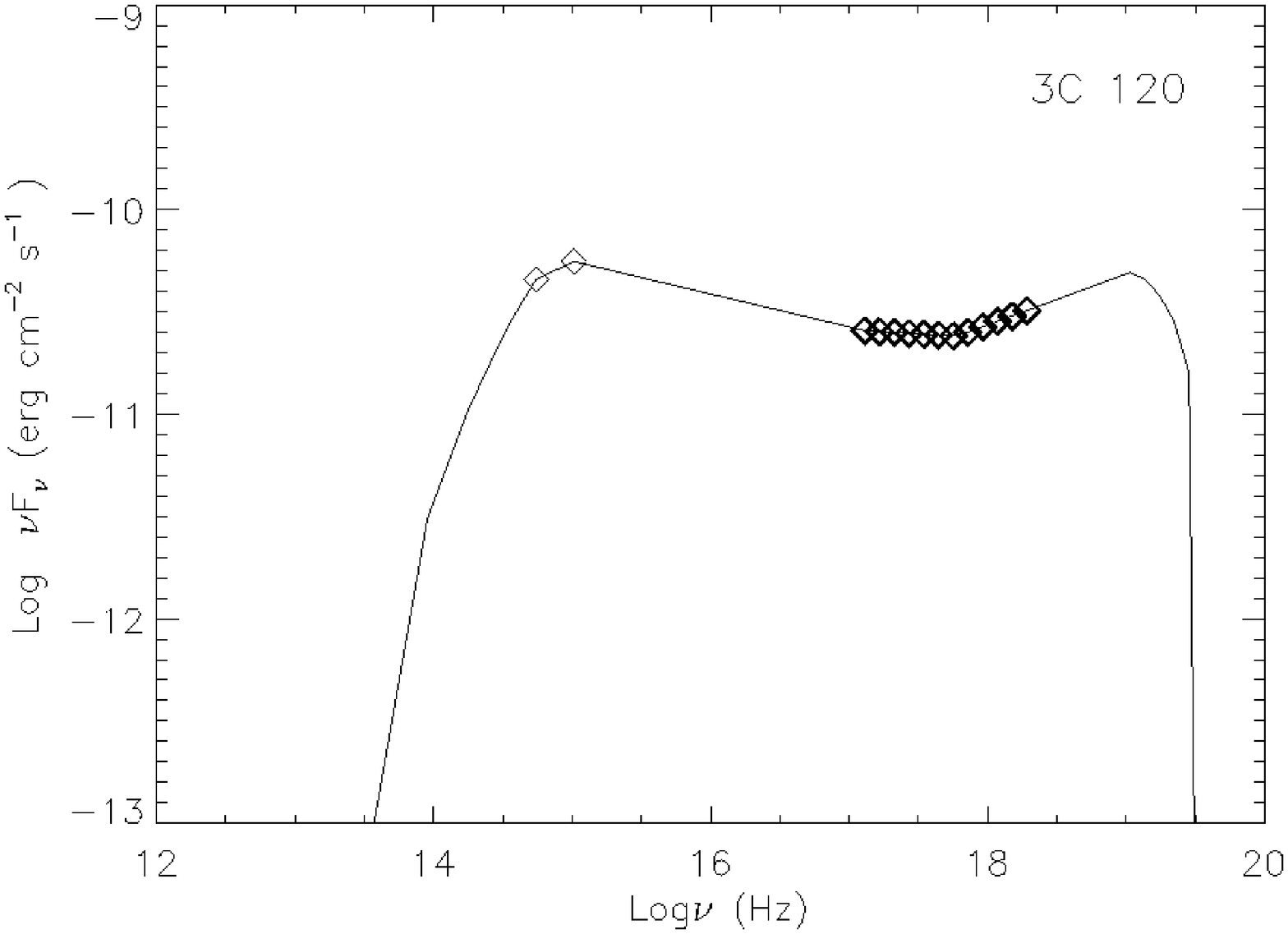,height=5cm,width=5cm, angle=90}
 \caption{Spectral energy distributions (SED) of 3C~390.3 and 3C~120, constructed from the \textit{XMM--Newton} data in the optical--UV--to--X--ray band, and from the standard AGN radio--IR continuum included in {\small CLOUDY}.}
   \label{fig1}
\end{center}
\end{figure}


\subsection{3C~390.3}
At first the continuum of both observations was modeled with a power law ($\Gamma \sim 1.9$) plus a neutral absorber ($N_{H}\sim 2.5\times10^{20} {\rm cm}^{-2}$) in addition to the Galactic one (C/d.o.f.=583/406). Although during the second pointing (Oct.~17) the source flux was about 14$\%$ lower, the spectral parameters are completely consistent.
 A careful inspection of the residuals revealed photon deficits in the regions of Ne{\small X} (12.134 \AA\ ), Fe{\small XX} (12.864 \AA\ ), O{\small VIII}~Ly$\alpha$ (18.969 \AA\ ) and N{\small VI} (24.898 \AA\ ). An example of absorbed structures is shown in Fig.~2.
In order to confirm the WA detection and constrain physical properties of the WA, the \textit{xabs} component was added to the continuum model. 
The column density of the ionized absorber N$_{H1}$, the ionization parameter $\xi$ and the velocity of the gas were let free to vary. The  velocity dispersion between different blend components is fixed to the default value v=100 km~s$^{-1}$.
The covering fraction parameter (f$_{cov}$) is fixed to the default value equal to 1. 
The fit improves with the addition of the warm absorber with respect to the power law alone, i.e. $\Delta$C=18 and 11 for a decrease by 3 in the number of degrees of freedom, corresponding to a significance $>$99.6$\%$ and $\sim$97$\%$, for the longest and shortest observations, respectively (see Table~2 and Fig.~3). \\
Although one--phase absorption model well describes the data, we also tested a stratified gas possibility. Actually, the inclusion of a second absorber, with a higher ionization parameter  (Log$\xi \approx 3$~erg~cm~s$^{-1}$) and column density ($N_{H2}\approx1.5\times10^{21}$~cm$^{-2}$) seems to better reproduce the shape of the most prominent features. However, the statistical improvement of the fit is not significant.
Finally, we note positive residuals around 23.5 \AA\ (observed--frame) in both data sets (Fig.~4). In agreement with Sambruna et al. (2009),  a narrow gaussian component at the theoretical wavelength of the O{\small VII} forbidden line (Table~3) is a good parametrization of this emission feature.

\begin{table*}
\centering
\caption{Best--fitting parameters for both 3C~390.3 observations. The observation date, photon index, source rest--frame neutral absorber column density (N$_{H}$), column density of the ionized absorber (N$_{H_1}$), ionization parameter (Log$\xi$), outflow velocity and $\Delta$C after the addition of the \textit{xabs} model to the absorbed power law are reported.}
\label{tab2}      
\centering          
\begin{tabular}{l c c c c c c}     
\\
\hline\hline\                    
         
Obs.      &$\Gamma$      &N$_{H}$                             &N$_{H1}$                         &Log$\xi$                   &v                                     &$\Delta$C  \\                                             
              &                       &(10$^{20}$cm$^{-2}$)        &(10$^{20}$cm$^{-2}$)      &(erg~cm~s$^{-1}$)  &(km~s$^{-1}$)                           & \\                         
\hline 
Oct.~8  &1.89$\pm$0.05 &2.5$\pm$0.2 &5.5$^{+2.0}_{-1.7}$ &2.08$^{+0.12}_{-0.07}$ &$<$600 &18  \\ 
\hline
Oct.~17 &1.96$\pm$0.02 &3.1$\pm$0.62 &3.7$^{+2.9}_{-1.9}$ &1.98$^{+0.2}_{-0.08}$ &$<$1000 &11\\
\hline
\end{tabular}
\end{table*}



\begin{figure}
\begin{center}
\epsfig{file=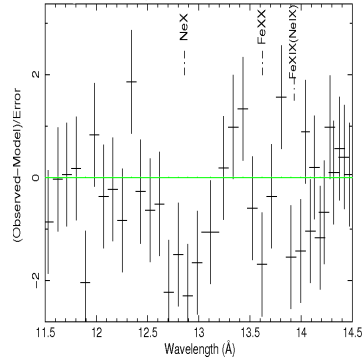,height=5cm,width=5cm, angle=-90}
 \caption{3C~390.3 residuals in a zoomed region around Ne{\small X}, Fe{\small XX} lines (observed--frame) after fitting the first observation (Oct.~8) with a power law plus a neutral absorber in addition to the Galactic one.}
   \label{fig2}
\end{center}
\end{figure}



\begin{figure}
\begin{center}
\epsfig{file=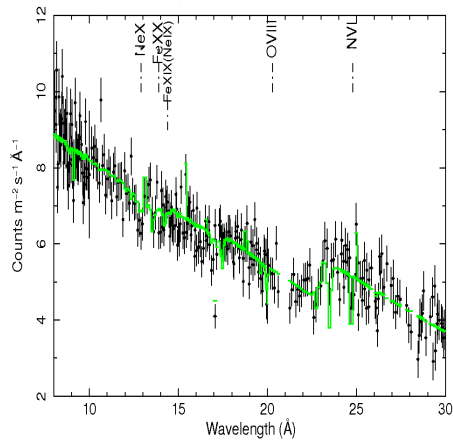,height=6cm,width=6cm, angle=-90}
 \caption{{\small SPEX} best--fit modeling of the RGS spectrum of 3C~390.3. The most prominent absorption lines are labelled.}
   \label{fig3}
\end{center}
\end{figure}


\begin{figure}
\begin{center}
\epsfig{file=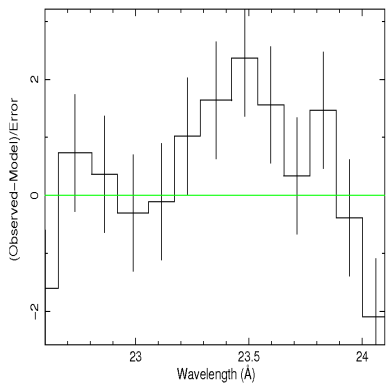,height=5cm,width=5cm, angle=-90}
\hspace{0.5cm}
 \caption{3C~390.3 residuals for the O{\small VII} forbidden line.}
   \label{fig4}
\end{center}
\end{figure}

\begin{table*}
\centering
\caption{3C~390.3 O{\small VII}(f) emission line parameters for the two epochs. The wavelength, flux and FWHM together with $\Delta$C after the addition of the line to the absorbed power law are reported.}
\label{tab3}      
\centering          
\begin{tabular}{l c c c c c}     
\\
\hline\hline\                    
         
Obs.      &Line        &$\lambda$      &Flux           &FWHM     &$\Delta$C \\                                             
             &               &(\AA\ )           &(10$^{-4}$~ph~cm$^{-2}$~s$^{-1}$)  &(\AA )  \\                         
\hline 
Oct.~8  &O{\small VII}(f) &22.101 &0.56$^{+0.68}_{-0.26}$ &0.51$^{+0.47}_{-0.23}$ &13\\
\hline
Oct.~17 &O{\small VII}(f)  &22.101 &0.64$^{+0.82}_{-0.24}$ &0.33$^{+0.21}_{-0.12}$  &17\\
\hline
\end{tabular}
\end{table*}

\subsection{3C~120}

The soft X--ray continuum of 3C~120 is very complex.
An RGS fit with a simple power law absorbed by two neutral absorbers is not a satisfying representation of the data (C/d.o.f.=803/328).  
The spectral shape of the continuum seems to be curved, as suggested by negative residuals around 23--26 \AA\ that is the region where O{\small I} edge is expected. We exclude that this spectral bending can be attributed to a warm absorber. Indeed the \textit{xabs} model is not statistically required by the data and left the residuals invariant. 
Following Ogle et al. (2005), the oxygen abundance of the second absorber was allowed to vary.  The fit greatly improves (C/d.o.f.=562/327) and the oxygen abundance drops to A$_{(O)}$=0.53$\pm$0.07. 
However an inspection of the residuals still reveals emission features in the range 17--20 \AA\ (Fig.~5).
These structures can be fitted with two gaussian lines at the wavelength corresponding to the rest--frame positions of Fe{\small XVII} and O{\small VIII}~Ly$\alpha$ (see Table~4), providing a further improvement of the fit (C/d.o.f.=449/323). 
To ascertain the nature of the emitting--line gas we tested both collisional (i) and photoionized (ii) scenarios.\\
(i) A single temperature collisional model, {\small CIE} (Collisional Ionization Equilibrium) in {\small SPEX}, was applied instead of two single lines. A thermal component with kT=0.37$^{+0.09}_{-0.06}$~keV can not completely take into account the emission features, infact residuals around the Fe{\small XVII} are still present. A second {\small CIE} component gives a poor fit C/d.o.f.=520/323, while a better modeling is provided by a non--equilibrium ionization jump model ({\small NEIJ}) in {\small SPEX} (C/d.o.f.=498/323). However, even in this case, there are still positive residuals.\\
(ii) If the emitting--line gas has a photoionized origin, the emission line at 16.892 \AA\ could be radiative recombination continuum (RRC) from O{\small VII} slightly redshifted with respect to the rest--frame of the galaxy (v$\sim$+2000~km~s$^{-1}$). We tested this hypothesis fitting this feature with the {\small RRC} model in {\small SPEX}. Since RRCs are narrow and prominent structures in photoionized plasmas, we assumed a typical electron temperature of kT=3 eV. We fit the RRC emission measure of O{\small VII} as a free parameter, however the fit is still not satisfying (C/d.o.f.=517/324).

\noindent
We conclude that our proposed models can not reproduce the data better than a single temperature gas plus a gaussian line, C/d.o.f.=440/323 (see Table~4), leaving the exact nature of the emitting--line gas still uncertain.
However we exclude that one, or two, plasma components in collisional ionization equilibrium, or even in non--equilibrium, are sufficient to reproduce the soft--excess in 3C~120.

\begin{figure}
\begin{center}
\epsfig{file=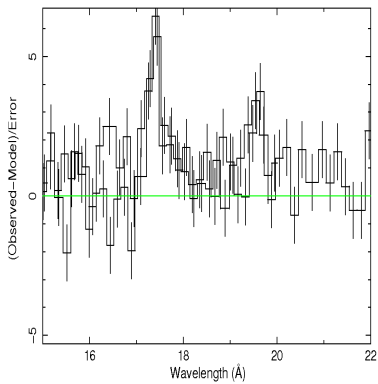,height=5cm,width=5cm, angle=-90}
 \caption{3C~120 residuals after fitting the RGS data with a power law plus two neutral absorbers, with the oxygen abundance of the second absorber free to vary.}
   \label{fig5}
\end{center}
\end{figure}

\begin{table*}
\centering
\caption{3C~120 fit parameters for the two models tested. For Model 1 C/d.o.f.=449/323, for Model 2 C/d.o.f.=440/323.}
\label{tab4}      
\centering          
\begin{tabular}{l c c}
\\
\hline\hline
                                       &Model~1                        &Model~2\\
\hline
$\Gamma$                     &2.32$\pm$0.04              &2.32$\pm$0.04\\

N$_{H}^{a}$                         &8.8$^{+0.5}_{-0.6}$        &8.8$\pm$0.7 \\

A$_{(O)}$                       &0.53$\pm$0.07                 & 0.55$\pm$0.06 \\

$\lambda$(\AA\ )         &16.892                             &16.892\\

Flux$^{b}$                      &0.66$^{+0.04}_{-0.2}$   &0.38$^{+0.07}_{-0.05}$    \\

FWHM(\AA\ )               &0.27$^{+0.13}_{-0.15}$  &0.13$^{+0.18}_{-0.07}$\\   

$\lambda$(\AA\ )          &18.969                            &--\\

Flux$^{a}$                       &0.18$^{+0.09}_{-0.08}$   &--\\

FWHM(\AA\ )               &0.13$^{+0.18}_{-0}$        &--\\

kT(keV)                              & --                                     &0.37$^{+0.09}_{-0.06}$\\
\hline
\\
\multicolumn{3}{l}{(a) $\times$ 10$^{20}$~cm$^{-2}$.}\\
\multicolumn{3}{l}{(b) in 10$^{-4}$~ph~cm$^{-2}$~s$^{-1}$.}\\
\end{tabular}
\end{table*}


\section{Discussion}
 \subsection{Warm Absorber  Physical Properties}
Table~5 lists, for each source, column density, ionization parameter, outflow velocity and minimum and maximum radii derived for the WAs of the studied BLRGs. For 3C~390.3 we consider the parameters obtained from the Oct.~8 observation for which the fit improvement  due to the inclusion of a warm absorber  is more significant. For 3C~445, the soft emitting gas properties, derived by R10, are also listed.
In order to establish the location of the emitting/absorbing region, the measured distances of the broad line region (BLR) and the torus are also reported.\\
The minimum distance of the WA (r$_{min}$) from the central engine is measured from:
\begin{equation}
r_{min}\geq \frac{2GM}{v_{out}^{2}}
\end{equation} 
where M is the black hole mass and assuming that the outflow must have a speed (v$_{out}$)  greater than or equal to the escape velocity. \\
The maximum distance  (r$_{max}$) can be estimated assuming that most of the mass of the absorber is concentrated in a thin layer $\Delta$R where $\Delta$R/R$\leq$1. 
The column density is a function of the density of the material n(R) at ionization parameter $\xi$, of its volume filling factor (here assumed equal to 1) and of $\Delta$R:
\begin{equation}
N_{H_{1}}\sim n(R)\Delta R C_{v}
\end{equation}
This combined with the expression of the ionization parameter gives 
\begin{equation}
\frac{\Delta R}{R}\sim \frac{\xi RN_{H_{1}}}{L_{ion}}
\end{equation} 
Therefore  if $\Delta$R/R$\leq$1
\begin{equation}
r_{max}\leq \frac{L_{ion}}{\xi N_{H_{1}}}
\end{equation} 

\noindent
The BLR and torus radii are calculated following the prescriptions of Ghisellini \& Tavecchio (2008), that simply assume that typical distances scale as the square root of the ionizing disk luminosity:
\begin{equation}
r_{BLR}=10^{17}L_{disk,45}^{1/2} {\rm cm}
\end{equation} 

\begin{equation}
r_{torus}=2.5\times10^{18}L_{disk,45}^{1/2} {\rm cm}
\end{equation} 

\noindent
Table~5 suggests two immediate considerations:\\
(i) depending on the line--of--sight, different features are revealed (see also Table~1). 
The absorption lines due to photoionization and photoexcitation processes are preferentially observed in sources seen at small \textit{i} (i.e. 3C~382 and 3C~390.3). On the contrary, the emission lines produced by the inverse processes are dominant in 3C~445, the only source with Seyfert 2 characteristics (and presumably with the larger jet inclination angle). We note that no WA could be detected in 3C~120, the BLRG in the sample with the smallest inclination angle (\textit{i}$<$21$^{\circ}$) and the only one with a $\gamma$--ray counterpart (Abdo et al. 2010b). However this source is quite complex. Indeed the detection of X--ray emission lines is  at odds with  the idea that the jet dominates  the soft X--ray emission; \\

\noindent
(ii) the location of the ionized outflow is not unique. In 3C~445 the WA was suggested by the detection of a strong edge around 7 keV.
The  deduced high velocity (v$_{out}$$\sim$10$^{4}$~km~s$^{-1}$), high column density (N$_{H}$$>$10$^{23}$~cm$^{-2}$) and low ionization parameter (Log$\xi$=1.4~erg~cm~s$^{-1}$) of the wind indicate a probable origin in the disk (R10). On the contrary, in 3C~382 and 3C~390.3, the absorption features, signatures of an ionized outflow, were found in the soft part of the grating spectra. The gas has different physical parameters and probably a different origin.
Indeed the column densities and ionization parameters vary in the range N$_{H}$$=$10$^{20-22}$~cm$^{-2}$ and log$\xi$=2--3~erg~cm~s$^{-1}$. Moreover the slower velocities, v$_{out}$$\sim$10$^{2-3}$~km~s$^{-1}$, constrain the location of such gas between the torus and the NLR, favoring the torus wind scenario (Krolik \& Kriss 2001; Blustin et al. 2005 hereafter B05). \\

\subsection{Warm  Absorber Energetics}
Warm absorber energetics are summarized in Table~6. The mass accretion rate $\dot{M}_{acc}$  (L$_{bol} \approx$L$_{acc}$=$\eta \dot{M}$c$^{2}$) was calculated for $\eta$=0.1.\\
The mass outflow rate estimates how much mass is carried out of the AGN through the wind, and can be expressed as (B05):
\begin{equation}
\dot{M}_{out}\sim \frac{1.23 m_{p} L_{ion} v_{out} C_{v}\Omega}{\xi}
\end{equation} 
We set the solid angle of the outflow $\Omega$=2.1, using the information that $\sim$33$\%$ of the Radio Galaxies belonging to the 3CR sample with z$\leq$1.5 are BLRGs (Buttiglione et al. 2009), and assuming that at least 50$\%$ of the objects possess an outflow as in Seyfert 1s.
The volume filling factor C$_{v}$, being unknown, was kept equal to 1, equivalent to assume an upper limit on $\dot{M}_{out}$.\\
The kinetic energy is the power released in the circumnuclear environment through the outflow and is expressed by:
\begin{equation}
\dot{E}_{out}=\frac{\dot{M}_{out}v_{out}^{2}}{2}
\end{equation} 
where v$_{out}$ is the blueshift velocity measured for the warm absorber. 

P$_{jet}$ is calculated according to the formula of Shankar et al. (2008) adapted from Willott et al. (1999): 
\begin{equation}
P_{jet}=3\times10^{45}f^{3/2}L^{6/7}_{151}  {\rm erg~s}^{-1} 
\end{equation}
L$_{151}$ is the observed radio luminosity in units of 10$^{28}$~W~Hz$^{-1}$~sr$^{-1}$ at 151 MHz. The factor \textit{f} accounts for systematic underestimates of the true jet power. The average value $<$f$>$=15  (Hardcastle et al. 2007) supports the picture of ``\textit{heavy}'' jets with a dominant protonic component. \\

\noindent
Looking at  Table~6, it appears immediately evident that the kinetic luminosity related to slow outflows is always a negligible fraction ($<<$1$\%$) of both bolometric luminosity and jet kinetic power and that the radiative power is generally larger than the kinetic one.  Shankar et al. (2008) proposed to directly link the kinetic power of the ejecting material (jets, winds) to the rest--mass energy of the accreting matter. If this is the case, P$_{jet}$ can also be expressed in terms of accretion energy: P$_{jet}$=$\eta_{jet} \dot{M}$c$^{2}$ (we do not take into account the wind, as its energetic contribution is not important). Then, the ratio between  P$_{jet}$ and L$_{bol}$  directly expresses  $\eta_{jet}/\eta$.  This efficiency ratio  is generally $< 1$, suggesting  that accretion power could be preferentially channeled in radiation rather than in jet kinetic power (at least in these sources). Assuming $\eta$ equal to 0.1 (typical value of standard accretion disks),  $\eta_{jet}$ ranges between 0.01--0.06. 
However if accretion disk winds are energetically important (see Section 4.3) the kinetic (jet + wind) power could compete with the radiative power.

\subsection{Comparison with type 1 Radio Quiet  AGNs}

Here we compare the X--ray properties of BLRGs studied in this work with a sample of type 1 RQ AGNs (Seyfert 1s, NLS1s, QSOs) having a good modeling of the WA. We chose the sample already studied by B05 with the addition of the Seyfert 1.2 IC4329a taken from McKernan et al. (2007). Among the sources exhibiting signatures of warm absorption as reported by McKernan et al., IC4329a is the only one not in common with B05. Very recently, a possible detection of a fast outflow in MR2251--178 has been proposed by Gofford et al. (2011) on the basis of  \textit{Suzaku} data. For this source, belonging to the B05 sample, we however prefer to consider only the WA physical parameters derived by the \textit{XMM--Newton}/RGS analysis  (Kaspi et al. 2004, B05). 
For the sake of consistency, when a source has a multi--phase outflow, only the higher phase, similar to BLRGs (Log$\xi$=2-2.9~erg~cm~s$^{-1}$), is considered. It is important to note that, in this case, the estimated mass outflow rates refer only to the higher phases, neglecting the contribute of mass outflow rates related to lower ionization parameters.\\
However, even considering the averaged warm absorber parameters the final result does not change. 
The mass outflow rates and the related kinetic powers of RQ objects in Table~4 of B05 are rescaled to a volume filling factor equal to 1 in order to match our assumption.  We keep the small (and of no consequence) difference between the solid angle of BLRGs ($\Omega$=2.1)  and Seyferts/QSOs  ($\Omega$=1.6), both estimated on the basis of similar considerations.  Fixing C$_v$=1, the RQ mass outflow rates are obviously shifted to larger values.\\

\noindent
As shown in Fig.~6 (\textit{upper panel}), all the sources have large (and implausible) mass outflow rates exceeding the mass accretion rates even more than one order of magnitude implying, also for RL sources, a non uniform distribution of the photoionized gas. 
We can deduce a BLRG  volume filling factor as small as $\sim$0.01, simply assuming that the same amount of matter is accreted  and ejected as wind from the nuclear region (i.e. $\dot{M}_{out}\sim \dot{M}_{acc}$ with $\eta$=0.1).\\
In both RL and RQ  AGNs,  the kinetic energy associated to the slow winds is negligible with respect to the radiative luminosity, even considering a uniform gas distribution (C$_v$=1) (Fig.~6 \textit{lower panel}). There are only three sources, PG~1211+143, PG~0844+349 and 3C~445, for which $\dot{E}_{out} \ge L_{bol}$($\approx$L$_{acc}$). In these objects the winds have velocities of several thousands km~s$^{-1}$, in spite of the presence/absence of a powerful jet, and are probably directly connected to the {accretion} disk (Pounds et al. 2003a,b).  These disk winds are  extremely energetic, unless they have small covering factors and/or are transient phenomena.
Indeed, in BLRGs  persistent disk outflows covering large solid angles could even compete with the  jets in trasferring momentum to the circumnuclear environment.  \\

\noindent
In order to further investigate the role of the relativistic jet, we explore a possible correlation between the mass outflow rate and the radio--loudness parameter (R). We adopt a radio--loudness:
\begin{equation}
R=Log[ \frac{\nu L_{\nu(5GHz)}}{L_{(2-10keV)}}]
\end{equation}
as proposed by Terashima \& Wilson (2003).
For RQ AGNs, we use the 2--10 keV X--ray luminosities as measured by Bianchi et al. (2009), while we refer to Torresi et al. (2010) and to this paper for the X--ray luminosities of 3C~382 and 3C~390.3, respectively.\\
The 5 GHz luminosities of RL and RQ sources are taken from Kellermann et al. (1969) and from Ulvestad et al. (1984), respectively.
For RQ objects, when the luminosity at 5 GHz is not available, we extrapolate it from the 1.4 GHz luminosity as reported in the NRAO/VLA Sky Survey catalogue (NVSS; Condon et al. 1998) by assuming $\alpha$=0 (Nagar, Wilson \& Falcke 2001). We notice that different assumptions on the radio spectral slope do not affect the radio--loudness estimation. Indeed even using $\alpha$=0.8, given the dispersion of spectral indexes for AGNs (Kukula et al. 1998), the estimated R values are completely consistent with those obtained for $\alpha$=0.\\
We consider BLRGs and type 1 AGNs having slow outflows (v$_{out}$=10$^{2-3}$~km~s$^{-1}$) and similar phases. Disk winds are not taken into account here because we have only three sources.
In Fig.~7 the mass outflow rates of both RL and RQ AGNs are plotted as a function of the radio--loudness. This plot suggests a possible difference between the two classes. Indeed the average $\dot{M}_{out}$ value of RQ objects is $\sim$4 M$_{\odot}$~yr$^{-1}$, much lower than that of BLRGs that is around 25 M$_{\odot}$~yr$^{-1}$.\\
When the Spearman $\rho$ test and the generalized Kendall $\tau$ test in the Astronomy Survival Analysis ({\small ASURV}) package (Feigelson et al. 1985; Isobe et al. 1986; Lavalley et al. 1992) are applied to data, a possible positive correlation between $\dot{M}_{out}$ and R is suggested. 
The resulting significance is $\alpha^{{\small ASURV}}_{\rho}$=0.05 and $\alpha^{{\small ASURV}}_{\tau}$=0.02, respectively. The correlation is assumed to be significant if equal or below 0.05.
This trend could simply reflect differences in the gas distribution, tending to preferentially clump when the system is less perturbed by the jet, given the dependence of $\dot{M}_{out}$ on the volume filling factor. Alternatively, if  the geometry of the gas is similar in both RQ and RL objects,  the observed correlation would indicate that larger amount of  mass escapes from the nuclear region  when the jet production mechanism is more efficient.
We notice that the ionization parameter, the outflow velocity and the ionizing luminosity are not correlated with R (significance of correlation 0.4, 0.8 and 0.2, respectively, according to the generalized Kendall $\tau$ test).


\begin{table*}
\caption{{\small For each source we report  the warm absorber/emitter parameters (LogN$_{H_{1}}$, Log$\xi$, v$_{out}$ (``+'' means the gas is redshifted, ``-'' is blueshifted)), together with the estimated minimum (r$_{min}$) and maximum (r$_{max}$) distances of the WA in pc, the distance of the BLR (r$_{BLR}$) and of the torus (r$_{torus}$) from the central engine, respectively.}}
\label{tab5}                
\begin{tabular}{l c c c c c c c}
\\
\hline                    
         
                                    &Log N$_{H_{1}}$    &Log $\xi$     &v$_{out}$ &r$_{min}$ &r$_{max}$ &r$_{BLR}$ &r$_{torus}$ \\                           
                                    &(cm$^{-2}$)      &(erg~cm~s$^{-1}$)     &(km~s$^{-1}$)   &(pc)    &(pc)     &(pc)   &(pc)    \\                
\hline 
\large\textit{3C~445} {\small (em)} &22.0   &3; 1.8    &+150$^{a}$;-430$^{b}$               & -           & $<$0.01$^{c}$; $\sim$0.1$^{d}$         &0.02    &0.45 \\
\hline
\large\textit{3C~445} {\small (wa)} &23.3    &1.4                &-10$^{4}$            &0.01              & -            & -          &-\\
\hline
\large\textit{3C~390.3} {\small (wa)} &20.7     &2.08        &$<$-600  &$\geq$9 &$\leq$450 &0.03 &0.7    \\
\hline
\large\textit{3C~382} {\small (wa)} &22.5 &2.69  &-1000  &$\geq$10 &$\leq$44 &0.05 &1.5 \\
\hline
\large\textit{3C~120}  & - &- &- &-&- &0.03 &0.8 \\
\hline
\\
\multicolumn{8}{l}{(a) Redshift velocity (compared to the systemic) of the emitter (R10).} \\
\multicolumn{8}{l}{(b) Outflow velocity determined by G07.} \\
\multicolumn{8}{l}{(c) Upper limit estimated from the definition of the ionization parameter of the emitting gas} \\
\multicolumn{8}{l}{R$^{2}$=L$_{ion}$/$\xi$n$_{e}$ (R10).}\\
\multicolumn{8}{l}{(d) Estimate from the measured line widths of the O{\small VII}--O{\small VIII} emission (R10).}\\
\end{tabular}
\end{table*}



\begin{table*}
\caption{{\small For each BLRG we give the mass outflow rate ($\dot{M}_{out}$), the kinetic luminosity of the outflow ($\dot{E}_{out}$), the bolometric luminosity (L$_{bol}$), the mass accretion rate ($\dot{M}_{acc}$) assuming $\eta$=0.1, the kinetic power of the jet (P$_{jet}$) and the jet extraction efficiency $\eta_{j}$=(P$_{jet}$/L$_{bol}$)$\eta$ .}}
\label{tab6}                
\begin{tabular}{l c c c c c c}
\\
\hline                    
         
             &Log$\dot{M}_{out}$                &Log$\dot{E}_{out}$     &Log L$_{bol}^{a}$     &$\dot{M}_{acc}$       &Log P$_{jet}$          &$\eta_{j}$\\                           
            & (M$_{\odot}$~yr$^{-1}$)  &(erg~s$^{-1}$)            &(erg~s$^{-1}$)       & (M$_{\odot}$~yr$^{-1}$)     &(erg~s$^{-1}$) &\\
\hline                     
\large\textit{3C~445} &2.90~C$_{v}$     &46.4             &45.10                             &0.2$^{b}$    &44.88  & 0.06\\
\hline
\large\textit{3C~390.3} &1.40~C$_{v}$         &41.66                   &45.65                             &0.77                   &45.12   &0.03\\
\hline
\large\textit{3C~382}     &1.41~C$_{v}$        &41.91                   &45.84                             &1.2                     &44.81   &0.01 \\
\hline
\large\textit{3C~120}     &-                             &-                          &45.34                              &0.37             &44.71   &0.02\\
\hline
\\
\multicolumn{6}{l}{(a) L$_{bol}$ directly estimated from the total SED of each source, except for 3C~445} \\
\multicolumn{6}{l}{taken from Marchesini et al. (2004).}\\
\multicolumn{6}{l}{(b) The value from R10 is rescaled assuming $\eta$=0.1.}\\
\end{tabular}
\end{table*}


\begin{figure}
\begin{center}
\epsfig{file=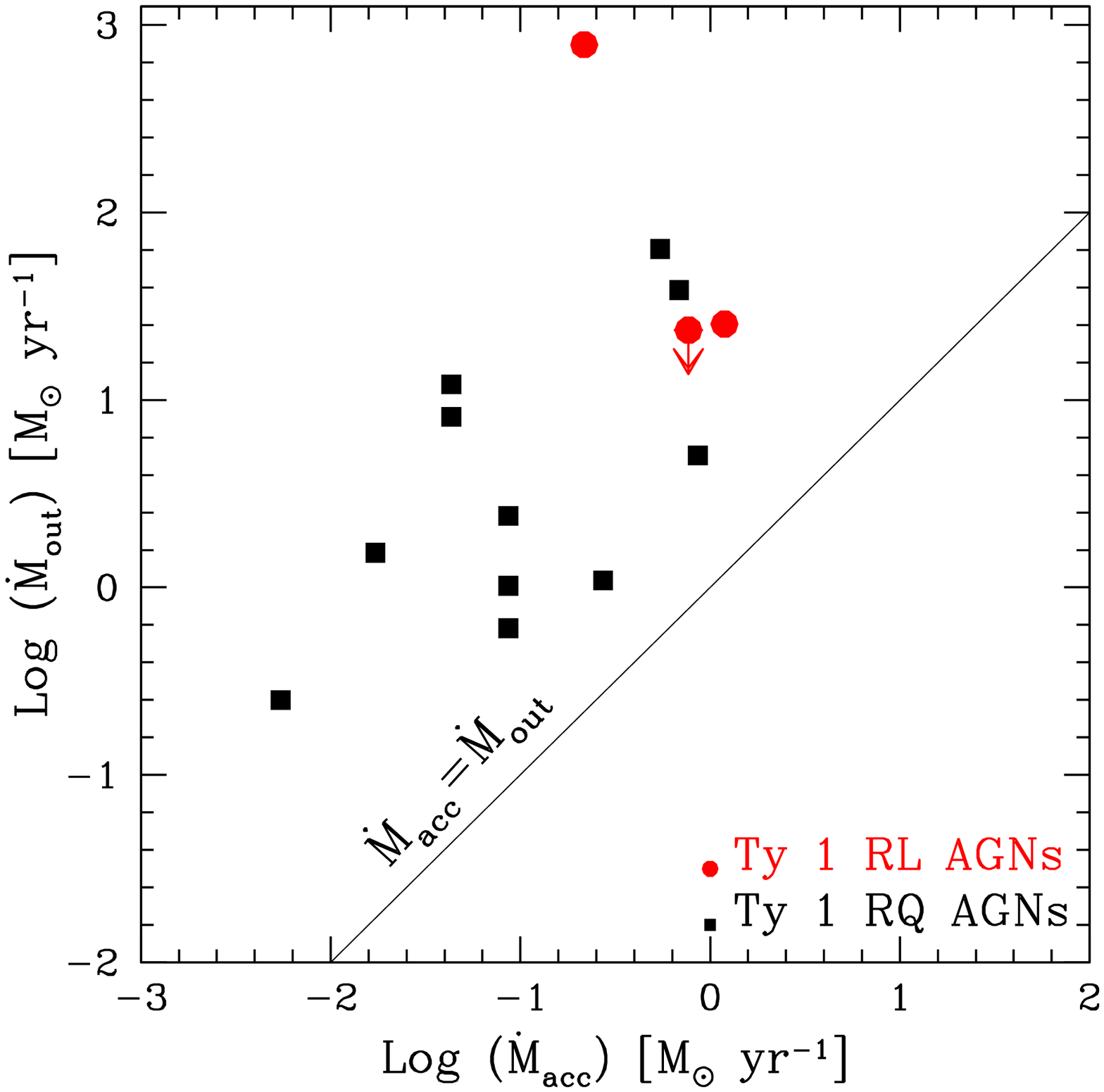,height=6cm,width=6cm, angle=0}
\epsfig{file=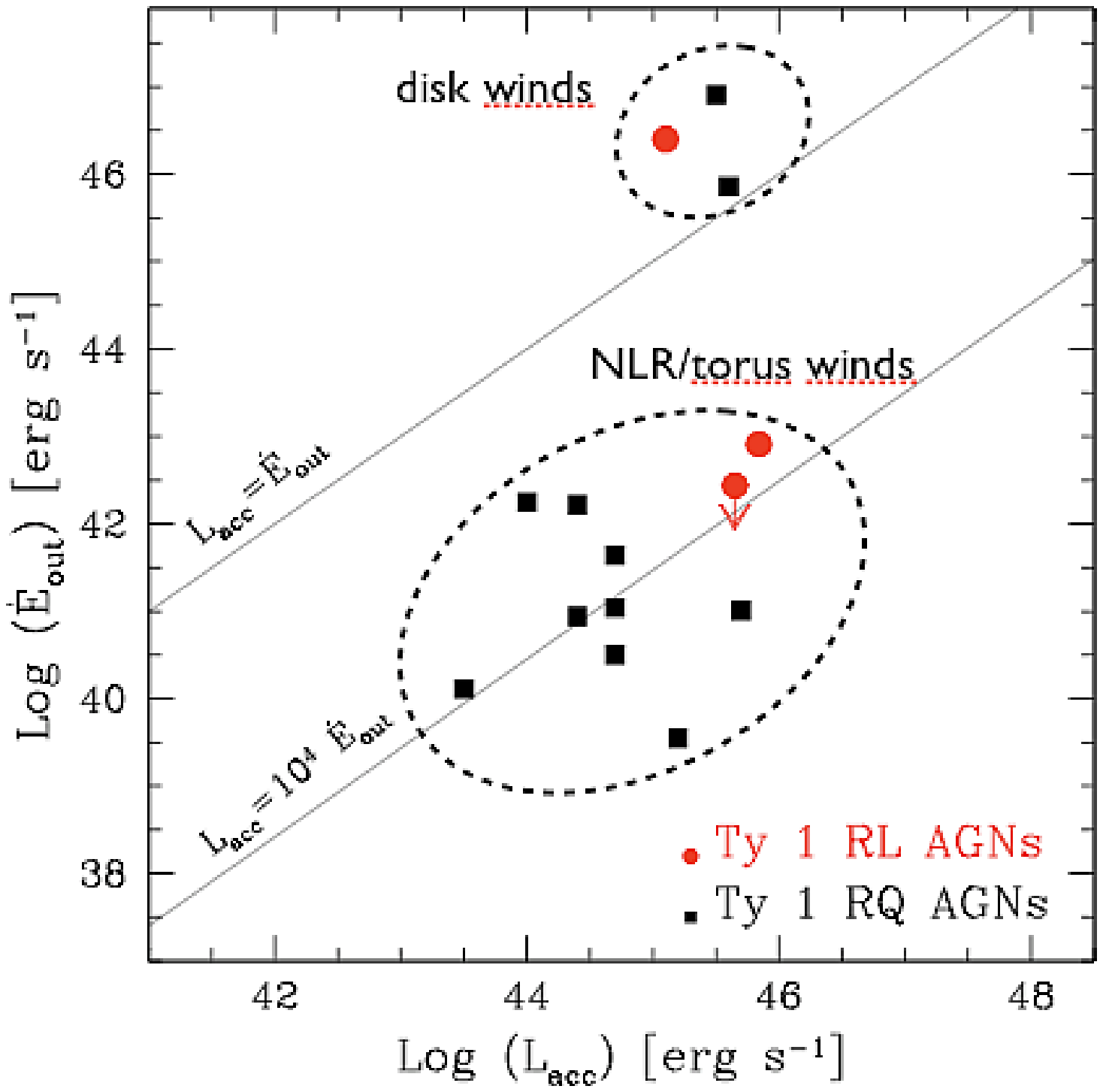,height=6cm,width=6cm, angle=0}
 \caption{\textit{Upper panel}: mass outflow rate ($\dot{M}_{out}$) plotted against the mass accretion rate ($\dot{M}_{acc}$). \textit{Lower panel}: kinetic luminosity associated to the outflow ($\dot{E}_{out}$) plotted against the accretion luminosity ($L_{acc}$). Type 1 RL AGNs (\textit{red circles}) are the BLRGs considered in this work; type 1 RQ AGNs (\textit{black squares}) refer to the Seyferts, NLS1s, and QSOs belonging to our reference sample.}
   \label{fig6}
\end{center}
\end{figure}

\begin{figure}
\begin{center}
\epsfig{file=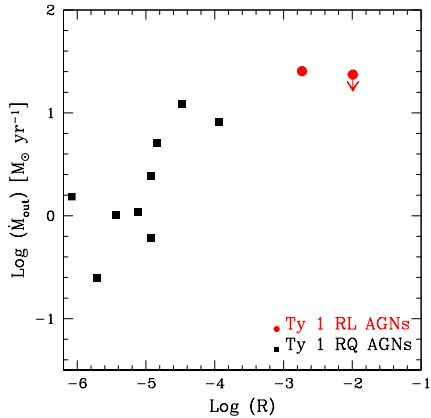,height=6cm,width=6cm, angle=0}
 \caption{Mass outflow rate plotted against the radio loudness (R=Log [$\nu$ L$_{\nu(5GHz)}$/L$_{(2-10keV)}]$). Type 1 RL AGNs (\textit{red circles}) are the BLRGs considered in this work; type 1 RQ AGNs (\textit{black squares}) refer to the Seyferts, NLS1s, and QSOs belonging to our reference sample.}
    \label{fig7}
\end{center}
\end{figure}

\begin{figure}
\begin{center}
\epsfig{file=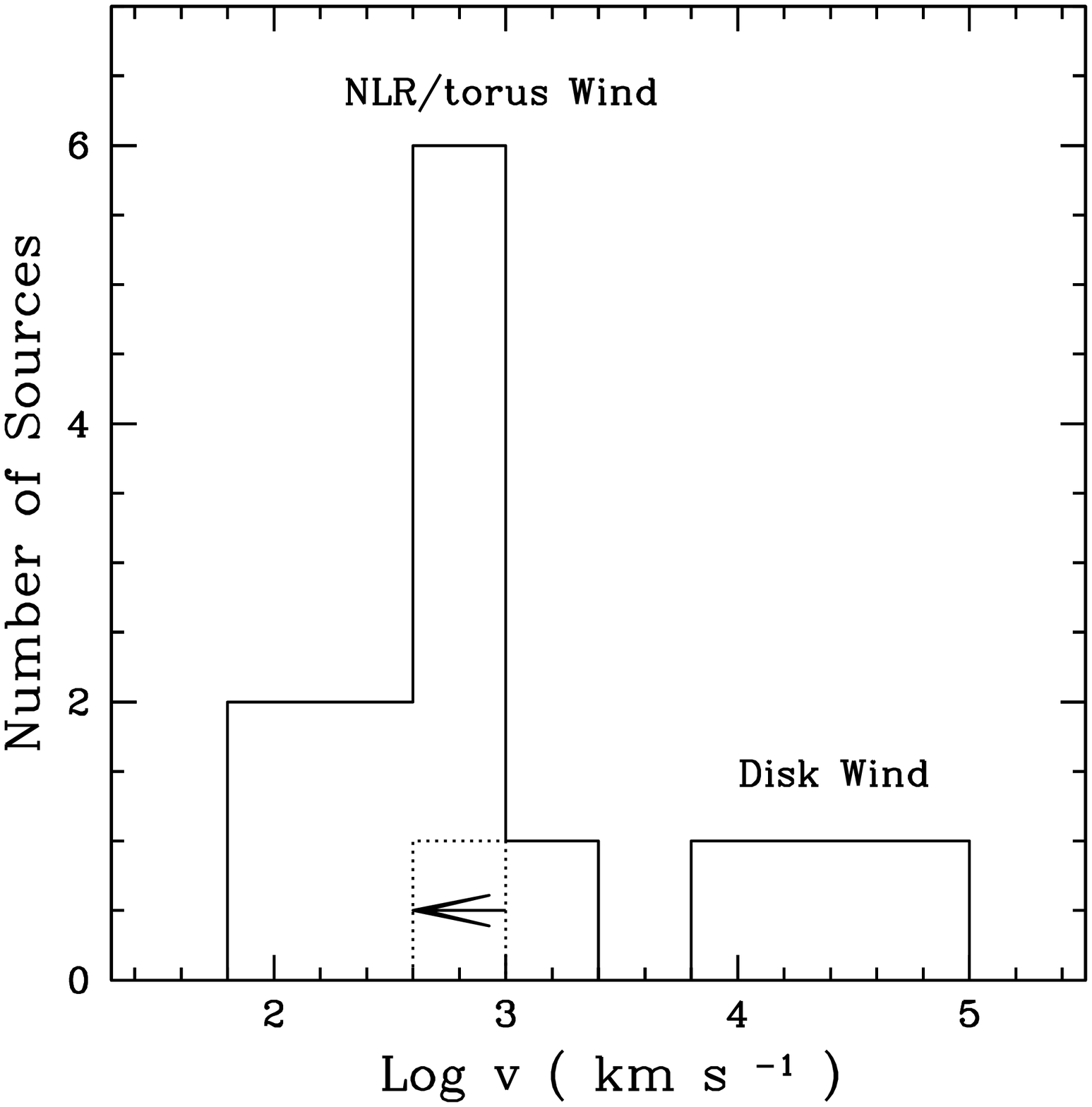,height=5cm,width=7.5cm, angle=0}
 \caption{Histogram showing the velocity distribution of RL and RQ outflows. The separation between NLR/torus winds and accretion disk winds is evident.}
   \label{fig8}
\end{center}
\end{figure}

\section{Conclusions}
In this work we report the detection of a warm absorber in the BLRG 3C~390.3. This is the third outflow observed after 3C~382 and 3C~445. On the contrary  3C~120, the BLRG with the smallest jet inclination angle (\textit{i}$ \leq$21$^{\circ}$) shows a complex soft X--ray spectrum only modified by a structured Galactic cold gas, without signatures of warm ionized absorption.\\
We discuss the physical and energetical properties of the warm absorbers found in three BLRGs, 3C~390.3, 3C~382 and 3C~445: 
\begin{itemize}
\item depending on the outflow velocities the absorbing gas has different locations: the NLR/torus for slower outflows (3C~390.3 and 3C~382), and the accretion disk for 3C~445. Interestingly, these winds have been detected in different spectral regions, i.e. the NLR/torus winds in the soft X--ray band (0.3--2 keV), while the accretion disk wind above 6 keV (R10);
\item the mass outflow rate ($\dot{M}_{out}$) of the absorbers is higher than the mass accretion rate ($\dot{M}_{acc}$), implying that we are overestimating the volume filling factor C$_{v}$, here assumed equal to 1 but probably much less than 1; 
\item even considering upper limits on $\dot{M}_{out}$, the kinetic luminosity associated with the slow outflows is always lower than the accretion luminosity and the jet kinetic power.
\end{itemize}
Aware of the scarcity of RL sources with WAs, we attempt a first comparison with a sample of type 1 RQ AGNs:
\begin{itemize}
\item fixing C$_{v}$=1, the mass outflow rates have implausibly large values in both RL and RQ AGNs, suggesting a clumpy gas configuration in all AGNs independently of their radio power;
\item in both Seyferts/QSOs and BLRGs, the kinetic luminosity related to slow outflows  is always negligible with respect to the accretion luminosity (and the jet kinetic power for RL AGNs);
\item fast accretion disk winds are observed in AGNs, independently of the RL or RQ classification (Fig.~8). The associated kinetic energies appear to be huge ($\dot{E}_{out} \geq$ L$_{acc}$) (unless these fast outflows have small covering factors and/or are transient phenomena);
\item although the RL and RQ WA properties are very similar (at least at zeroth order), the mass outflow rate ($\dot{M}_{out}$) and the radio--loudness parameter (R) seem to be correlated (Fig.~7) indicating a possible effect of a strong radio source on the outflowing winds.
Considering that  $\dot{M}_{out}$  depends on the covering factor, this result could simply indicate a different gas distribution in RL and RQ sources, with the WA  being  clumpier  in absence of a strong radio source. Alternatively, if the gas distribution is the same, the correlation could suggest that powerful jets favor the escape of more massive (but not necessarily more energetic) winds.

\end{itemize} 

\section*{Acknowledgments}
The anonymous referee is gratefully acknowledged for thoughtful comments on the manuscript. We warmly thank Jelle Kaastra for accurate and constructive comments and suggestions. We thank Kazushi Iwasawa for valuable and useful discussions. ET is grateful to Adriano de Rosa for providing constant technical assistance. 
\textit{XMM--Newton} is an ESA science mission with instruments and contributions directly funded by ESA Member States and NASA.
ET acknowledges the support of the Italian Space Agency (contract ASI/INAF I/009/10/0 and ASI/GLAST I/017/07/0).

{}


\newpage

\begin{thebibliography}{}

\bibitem[Abdo et al. 2010b]{magn} Abdo, A.A., et al. 2010b, ApJ, 720, 912

\bibitem[Alef et al. 1996]{alef96} Alef, W., Wu, S. Y., Preuss, E., Kellermann, K. I., Qiu, Y. H. 1996, A\&A, 308, 376

\bibitem[Anders \& Grevesse 1989]{anders89} Anders, E., Grevesse, N. 1989, GeCoA, 53, 197

\bibitem[Arshakian et al. 2010]{L151_120} Arshakian, T.G., Torrealba, J., Chavushyan, V.H., Ros, E., et al. 2010, A\&A, 520, A62

\bibitem[Ballantyne et al. 2004]{ballantyne04} Ballantyne, D.R., Fabian, A.C., Iwasawa, K. 2004, MNRAS, 354, 839

\bibitem[Ballantyne 2005]{ballantyne05} Ballantyne, D.R. 2005, MNRAS, 362, 1183

\bibitem[Balmaverde Capetti \& Grandi 2006]{Balmaverde06} Balmaverde, B., Capetti, A., Grandi, P. 2006, A\&A, 451, 35

\bibitem[Belsole et al. 2006]{belsole06} Belsole, E., Worrall, D.M., Hardcastle, M.J., 2006, MNRAS, 366, 339

\bibitem[Bianchi et al. 2009]{caixa} Bianchi, S., Guainazzi, M., Matt, G., Fonseca Bonilla, N., Ponti, G. 2009, A\&A, 495, 421

\bibitem[Blacket al. 1992]{black92} Black, A.R.S., Baum, S.A., Leahy, J.P., Perley, R.A., Riley, J.M., Scheuer, P.A.G. 1992, MNRAS, 256, 186

\bibitem[Blustin et al. 2002]{blustin3783} Blustin, A.J., Branduardi-Raymont, G., Behar, E., et al. 2002, A\&A, 392, 453 

\bibitem[Blustin et al. 2005]{blustin05} Blustin, A.J., Page, M.J., Fuerst, S.V., et al. 2005, A\&A, 431, 111 (B05)

\bibitem[Blustin \& Fabian 2009]{blustin09} Blustin, A.J., Fabian, A.C. 2009, MNRAS, 396, 1732

\bibitem[Bohlin et al.(1978)]{bohlin78} Bohlin, R.~C., Savage, B.~D., \& Drake, J.~F. 1978, ApJ, 224, 132 

\bibitem[Braito et al. 2011]{braito11} Braito, V., Reeves, J.N., Sambruna, R.M., Gofford, J. 2011, MNRAS, 414, 2739

\bibitem[Burbidge 1967]{burbidge} Burbidge, E.M., 1967, ApJ, 149, L51

\bibitem[Buttiglione et al. 2009]{buttiglione09} Buttiglione, S., Capetti, A., Celotti, A., Axon, D.J., Chiaberge, M., et al. 2009, A\&A, 495, 1033 

\bibitem[Cardelli et al.1989]{cardelli} Cardelli, J.A., Klayton, G.C., Mathis, J.S.,  1989, ApJ, 345, 254

\bibitem[Cleary et al. 2007]{cleary07} Cleary, K., Lawrence, C.R., Marshall, J.A., Hao, L., Meier, D. 2007, ApJ, 660, 117

\bibitem[Condon et al. 1998]{nvss} Condon, J.J., Cotton, W.D., Greisen, E.W., et al. 1998, AJ, 115, 1963

\bibitem[Crenshaw et al. 2003]{crenshaw03} Crenshaw, D.M., Kraemer, S.B., \& George, I.M. 2003, ARA\&A, 41, 117

\bibitem[den Herder et al. 2001]{denherder01} den Herder, J. W., Brinkman, A. C., Kahn, S. M., Branduardi-Raymont, G., et al. 2001, A\&A, 365, 7 

\bibitem[Eracleous \& Halpern(1994)]{1994ApJS...90..1} Eracleous, M., \& Halpern, J.~P. 1994, ApJS, 90, 1

\bibitem[Eracleous et al. 2000]{eracleous00} Eracleous, M., Sambruna, R.M., Mushotzky, R.F. 2000, ApJ, 537, 654

\bibitem[Evans et al. 2010]{evans10} Evans, D. A., Reeves, J. N., Hardcastle, M. J., Kraft, R. P., Lee, J. C., Virani, S. N. 2010, ApJ, 710, 859

\bibitem[Feigelson \& Nulsen 1985]{asurv1} Feigelson, E.D. \& Nulsen, P.I. 1985, ApJ, 293, 192 

\bibitem[Ferland et al.(1998)]{ferland98} Ferland, G.~J., Korista, K.~T., Verner, D.~A., Ferguson, J.~W., Kingdon, J.~B.,  \& Verner, E.~M. 1998, PASP, 110, 761 

\bibitem[Ghisellini \& Tavecchio 2008]{ghisellinietavecchio} Ghisellini, G. \& Tavecchio, F. 2008, MNRAS, 387, 1669

\bibitem[Ghosh 1991]{ghosh91} Ghosh, K. K., Sondararajaperumal, S. 1991, AJ, 102, 1298

\bibitem[Giovannini et al. 2001]{giova01} Giovannini, G., Cotton, W. D., Feretti, L., Lara, L., Venturi, T. 2001, ApJ, 552, 508

\bibitem[Gliozzi et al. 2003]{gliozzi03} Gliozzi, M., Sambruna, R.M., Eracleous, M. 2003, ApJ, 584, 176

\bibitem[Gliozzi et al.(2007)]{gliozzi3c382} Gliozzi, M., Sambruna, R. M., Eracleous, M., \& Yaqoob, T. 2007, ApJ, 664, 88

\bibitem[Gomez et al.(2001)]{gomez2001} Gomez, J., Marscher, A.P., Alberdi, A., et al. 2001, ApJ, 561, L161

\bibitem[Grandi et al. 1997]{grandi97} Grandi, P., Sambruna, R.M., Maraschi, L., Matt, G., Urry, C.M., Mushotzky, R.F. 1997, ApJ, 487, 636

\bibitem[Grandi et al. 1999]{grandi99}Grandi, P., Guainazzi, M., Haardt, F., Maraschi, L., Massaro, E., Matt, G., Piro, L., Urry, C. M. 1999, A\&A, 343, 33

\bibitem[Grandi et al. 2001]{grandi01}  Grandi, P., Maraschi, L., Urry, C.M., \& Matt, G. 2001, ApJ, 556, 35

\bibitem[Grandi et al.(2006)]{2006ApJ...642..113G} Grandi, P., Malaguti, G., \& Fiocchi, M. 2006, ApJ, 642, 113 

\bibitem[Grandi et al. 2007]{paola3c445} Grandi, P., Guainazzi, M., Cappi, M., Ponti, G. 2007, MNRAS, 381, L21 (G07)

\bibitem[Grevesse \& Sauval 1998]{solarabund} Grevesse, N., \& Sauval, A. J. 1998, Space Sci. Rev., 85, 161

\bibitem[Guainazzi 2010]{guainazzi 2010} Guainazzi, M. 2010, ``XMM-EPIC status of calibration and data analysis'', XMM--SOCCAL--TN-0018

\bibitem[Hardcastle et al. 1998]{hardcastle98} Hardcastle, M.J., Alexander, P., Pooley, G.G., \& Riley, J.M. 1998, MNRAS, 296, 445

\bibitem[Hardcastle \& Worrall 1999]{hardcastle99} Hardcastle, M.J., \& Worrall, D.M. 1999, MNRAS, 309, 969

\bibitem[Hardcastle et al. 2007]{hardcastle07} Hardcastle, M.J., Evans, D.A., Croston, J.H. 2007, MNRAS, 376, 1849

\bibitem[Harris et al. 2004]{harris04} Harris, D.E., Mossman, A.E., Walker, R.C. 2004, ApJ, 615, 161

\bibitem[Heckman et al. 1986]{heckman86} Heckman, T.M., Smith, E.P., Baum, S.A. et al. 1986, ApJ, 311, 526

\bibitem[Hewitt \& Burbidge 1991]{hewiti91} Hewitt, A., Burbidge, G. 1991, ApJS, 75, 297

\bibitem[Isobe et al. 1986]{asurv2} Isobe, T., Feigelson, E.D., Nelson, P.I. 1986, ApJ, 306, 490

\bibitem[Kaastra, Mewe and Nieuwenhuijzen 1996]{spex} Kaastra J.S., Mewe R., Nieuwenhuijzen H. 1996, in Yamashita K., Watanabe T., eds, UV and X-Ray Spectroscopy of Astrophysical and Laboratory
Plasmas. Universal Academy Press, Tokyo, p. 411

\bibitem[Kaastra et al. 2000]{k00} Kaastra, J.S., Mewe, R., Liedhal, D.A., Komossa, S., \& Brinkman, A.C. 2000, A\&A, 354, 83

\bibitem[Kalberla et al. 2005]{lab} Kalberla, P.M.W., Burton, W.B., Hartmann, Dap, Arnal, E.M., Bajaja, E., Morras, R., \& P\"oppel, W.G.L. 2005, A\&A, 440, 775 

\bibitem[Kaspi et al. 2002]{kaspi02} Kaspi, S., Brandt, W.N., George, I.M., Netzer, H., Crenshaw, D.M., et al. 2002, ApJ, 574, 643

\bibitem[Kaspi et al. 2004]{kaspi04} Kaspi, S., Netzer, H., Chelouche, D., George, I.M., Nandra, K., Turner, T.J. 2004, ApJ, 611, 68

\bibitem[Kataoka et al. 2007]{kataoka07} Kataoka, J., Reeves, J.N., Iwasawa, K., et al. 2007, PASJ, 59, 279

\bibitem[Kellermann et al. 1969]{kellermann69} Kellermann, K.I., Pauliny--Toth, I.I.K., Williams, P.J.S. 1969, ApJ, 157, 1

\bibitem[Kinkhabwala et al. 2002]{kinkhabwala02} Kinkhabwala A., Sako M., Behar E., et al. 2002, ApJ, 575, 732 

\bibitem[Komatsu et al. 2009]{cosmo09} Komatsu, E., et al. 2009, ApJS, 180, 330 

\bibitem[Krolik \& Kriss]{kk01} Krolik, J.H., \& Kriss, G.A. 2001, ApJ, 561, 684

\bibitem[Krongold et al. 2007]{krongold07} Krongold, Y., Nicastro, F., Elvis, M., et al. 2007, ApJ, 659, 1022 

\bibitem[Kruper et al. 1990]{kruper90} Kruper, J. S., Canizares, C. R., Urry, C. M. 1990, ApJS, 74, 347

\bibitem[Lavalley et al. 1992]{asurv3} Lavalley, M.P., Isobe, T., Feigelson, E.D. 1992, Bull. Am. Astron. Soc., 24, 839

\bibitem[Leighly \& O'Brien 1997]{leighly97} Leighly, K. M., O'Brien, P. T. 1997, ApJ, 481, 15

\bibitem[Lister et al. 2009]{lister09} Lister, M. L., Cohen, M. H., Homan, D. C., Kadler, M., Kellermann, K. I., Kovalev, Y. Y., Ros, E., Savolainen, T., Zensus, J. A. 2009, AJ, 138, 1874

\bibitem[Maraschi et al. 1991]{maraschi91} Maraschi, L., Chiappetti, L., Falomo, R., et al. 1991, ApJ, 368, 138

\bibitem[marchesini]{marchesini} Marchesini, D., Celotti, A., Ferrarese, L. 2004, MNRAS, 351, 733

\bibitem[Marzke et al. 1996]{marzke96} Marzke, R.O., Huchra, J.P., Geller, M.J. 1996, AJ, 112, 1803

\bibitem[Mason et al. 2001]{OM} Mason, K. O., Breeveld, A., Much, R., et al. 2001, A\&A, 365, 36  

\bibitem[Mathews \& Ferland(1987)]{mf87} Mathews, W.~G., \& Ferland, G.~J. 1987, ApJ, 323, 456 

\bibitem[McKernan et al. 2003]{mckernan03} McKernan, B., Yaqoob, T., Mushotzki, R.F., Ian, M.G., Turner, T.J. 2003, ApJ, 598, L83

\bibitem[McKernan et al. 2007]{mckernan07} McKernan, B., Yaqoob, T., Reynolds, C.S. 2007, MNRAS, 379, 1359

\bibitem[Nagar et al. 2001]{nagar01} Nagar, N.M., Wilson, A.S., \& Falcke, H. 2001, ApJ, 559, L87

\bibitem[Nicastro et al. 1999]{nicastro351} Nicastro, F., Fiore, F., Perola, G. C., Elvis, M., O'Dea, C.P. 1998, PASP, 110, 493

\bibitem[Ogle et al. 2005]{ogle05} Ogle, P.M., Davis, S.W., Antonucci, R.R.J., et al. 2005, ApJ, 618, 139

\bibitem[Pearson \& Readhead 1988]{pearsonreadhead} Pearson, T.J., Readhead, A.C.S. 1988, ApJ, 328, 114

\bibitem[Peterson et al. 2004]{peterson04} Peterson, B.M., et al. 2004, ApJ, 613, 682

\bibitem[Piconcelli et al. 2008]{3c234} Piconcelli, E.,  Bianchi, S., Miniutti, G., Fiore, F., Guainazzi, M., Jimenez-Bailon, E., Matt, G. 2008, A\&A, 480, 671

\bibitem[Pounds et al. 2003]{pounds03a} Pounds, K.A., King, A.R., Page, K.L., O'Brien, P.T. 2003a, MNRAS, 346, 1025

\bibitem[Pounds et al. 2003]{pounds03b} Pounds, K.A., Reeves, J.N., King, A.R.  2003b, MNRAS, 345, 705

\bibitem[Prieto 2000]{prieto00} Prieto, M.A. 2000, MNRAS, 316, 442

\bibitem[Reeves et al. 2003]{r03} Reeves, J.N., O' Brien, P.T., Ward, M.J. 2003, ApJ, 593, 65

\bibitem[Reeves et al. 2009]{reeves09} Reeves, J. N., Sambruna, R.M., Braito, V., Eracleous, M. 2009, ApJ, 702, 187

\bibitem[Reeves et al. 2010]{reeves10} Reeves, J. N., Gofford, J., Braito, V., Sambruna, R. 2010, ApJ, 725, 803 (R10)

\bibitem[Reynolds 1997]{reynolds97} Reynolds, C. 1997, MNRAS, 286, 513

\bibitem[Sambruna et al. 1999]{sambruna99} Sambruna R.M., Eracleous M., Mushotzky R.F. 1999, ApJ, 526, 60

\bibitem[Sambruna et al. 2007]{sambruna07} Sambruna, R., Reeves, J. N., Braito, V.  2007, ApJ, 665, 1030

\bibitem[Sambruna et al. 2009]{sambruna09} Sambruna, R. M., Reeves, J. N., Braito, V., Lewis, K. T., Eracleous, M., et al. 2009, ApJ, 700, 1473 

\bibitem[Seielstad et al. 1979]{seielstad79} Seielstad, G.A., Cohen, M.H., Linfield, R.P., et al. 1979, ApJ, 229, 53

\bibitem[Shankar et al. (2008)]{shankar08} Shankar, F., Cavaliere, A., Cirasuolo, M., \& Maraschi, L. 2008, ApJ, 676, 131

\bibitem[Stru\"der et al. 2001]{pn} Stru\"der, L., Briel, U., Dennerl, K., et al. 2001, A\&A, 365, 18

\bibitem[Terashima \& Wilson 2003]{radio_loudness} Terashima, Y., \& Wilson, A.S. 2003, ApJ, 583, 145

\bibitem[Torresi et al. 2009]{torresi09} Torresi, E., Grandi, P., Guainazzi, M., Palumbo, G.G.C., Ponti, G., Bianchi, S.  2009, A\&A, 498, 61 

\bibitem[Torresi et al. 2010]{torresi10} Torresi, E., Grandi, P., Longinotti, A. L., Guainazzi, M., Palumbo, G. G. C., Tombesi, F., Nucita, A. 2010, MNRAS, 401L, 10

\bibitem[Ulvestad et al. 1984]{ulvestad84} Ulvestad, J.S., Wilson, A.S. 1984, ApJ, 285, 439

\bibitem[Urry \& Padovani 1995]{urry_pad95} Urry, C.M., \& Padovani, P. 1995, PASP, 107, 803

\bibitem[Walker et al. 1987]{Walker87} Walker, R.C., Benson, J.M., Unwin, S.C. 1987, ApJ, 316, 546

\bibitem[Wamsteket et al. 1997]{Wamsteker97} Wamsteker, W., Wang, T.-G., Schartel, N., Vio, R. 1997, MNRAS, 288, 225

\bibitem[Willott et al. (1999)]{willott99} Willott, C.J., Rawlings, S., Blundell, K.M., \& Lacy, M. 1999, MNRAS, 309, 1017

\bibitem[Yee \& Oke 1981]{yee81} Yee, H.K.C., Oke, J.B. 1981, ApJ, 248, 472

\bibitem[Yaqoob \& Padmanabhan]{yaq04} Yaqoob, T, Padmanabhan, U. 2004, ApJ, 604, 63 

\bibitem[Zdziarski \& Grandi 2001]{zdz01} Zdziarski, A.A., Grandi, P. 2001, ApJ, 551, 186

\bibitem[Zheng 1996]{zheng96}Zheng, W. 1996, AJ, 111, 1498

\end{thebibliography}
\end{document}